\documentstyle[12pt,a4]{article}
\input{epsf.tex}
\begin{document}
\begin{center}
{\bf 
CALCULATION OF THE TOTAL GAMMA-SPECTRA OF THE FAST NEUTRONS CAPTURE IN THE ISOTOPES
$^{117,119} $Sn FOR THE DIFFERENT PARAMETERS OF CASCADE GAMMA-DECAY}
\end{center}

\begin{center}
{\  A.M. Sukhovoj, V.A. Khitrov}\\
\end{center}\begin{center}
{\it Frank Laboratory of Neutron Physics, Joint Institute
for Nuclear Research, 141980, Dubna, Russia}\\
\end{center}

The gamma-spectra  were calculated for the set of different level densities and
radiative strength functions. The sufficiently precise 
reproduction of the experiment is impossible without taking into account
the influence of the process of the nucleons Cooper pairs breaking on 
any nuclei cascade gamma-decay parameters.

\section{Introduction}

The direct determination of density $\rho$ of excitation levels
(number of levels of nucleus in the unit interval of excitation energy)
for the larger part of stable and long-life radioactive target nuclei is
impossible.
This assertion relates also to the radiative strength functions
\begin{equation} k=\Gamma_{\lambda i}/(E_{\gamma}^3\times
A^{2/3}\times D_{\lambda})
\end{equation}
exciting their primary dipole electrical and magnetic
gamma-transitions of level of nucleus decaying the excited
in the nuclear reaction. Extraction of the parameters of nucleus
in question in this situation can be executed by only their
fitting to the optimum values, reproducing the experimental
spectra and cross section with the minimum standard deviation
measured in the nuclear reactions.

This inverse problem of mathematical analysis by its nature
is principally ambiguous. Moreover, systems of equations,
connecting the number of excited levels and probability of
the emission of nuclear products are  usually assigned
within the framework of some ideas about the mechanism of
nuclear reaction and factors determining the dynamics of
the studied process.

Thus, for example, the description of the cascade gamma-decay
of neutron resonance, is impossible at present without the
introduction of some a priori ideas. In particular, within
the framework of the ideas about this process  following
potential possibilities are not taken into consideration:
\begin{itemize}
\item the presence of the possible strong dependence of neutron widths
$\Gamma_n$ on the structure of the wave function of
resonances (and of the excessive error in determination
of their density $D_{\lambda}^{-1}$ by the neutron time-of-flight method),

\item the analogous dependence of the partial radiative widths of
primary gamma-transitions $\Gamma_{\lambda i}$
on the structure of the level excited
by them, evidently overstepping the limits of
the expected Porter-Thomas fluctuations.
\end{itemize}

Potencial possibility of existence of the enumerated effects and
their significant influence on the process of cascade gamma-decay
directly follows from the results [1] of model approximation
of the level density, extracted from the reaction $(n,2\gamma)$
and the comparison [2,3] of the average values of the sums of
radiative strength functions with their models [4,5] most often
used in practice.
Thus, the possible break of sequential Cooper pair [1] with the
excitation energy in the region of the neutron-binding energy
can change values of few quasi-particle components in the wave
function and thus - change [6] values of $\Gamma_n$.
This possibility directly follows from the results, presented in [1].
Whether this possibility is realizable in principle, to what
degree the process of fragmentation of nuclear states mixes up
components of different types in the wave functions of the levels
in the whole region $E_{ex} \leq B_n$ and at the noticeably higher
excitation energies - neither the experiment, nor the theory
can answer this at the present.

In particular it is not possible to obtain the realistic estimation
of the part of the unobservable levels, which according to the
values $J^{\pi}$ could be excited them as s-resonances.
This problem is very essential, since the density of neutron
resonances in practice in any experiment to determine this
parameter for the excitation lower than $B_n$ is used
to standardize its relative values.
As a consequence of the above mentioned facts, the measured
experimentally in different procedures [2,3,7,8] level densities
and the radiative strength functions of primary gamma-transitions
can have an unknown systematic error, the value of which directly
depends on a systematic error in the conventional values
$D_{\lambda}$ of the spacing between the neutron resonances.
And the obtained ideas about these values and properties of
nucleus can be erroneous to a greater or lesser extent.
However, if we add the fundamental incompatibility of the data
about the level density between the results of applying the
procedures [7,8] on the one hand, and [2,3] on the other hand,
than the need for a maximally possible verification of $\rho$
and $k$ determined from indirect experiments becomes obvious.

\section{Possibilities and the specific character of the
verification of the experimentally determined values
of $\rho$ and $k$}

The verification of the indicated parameters of nucleus can
be partially executed by the calculation of total gamma-spectra
for different sets of $\rho$ and $k$ with their subsequent
comparison with the experiment. This calculation was carried
out by different authors repeatedly [9,10], but,
as a rule, without taken into account of:
\begin{itemize}

\item
the nonconformity of the model assigned ones and real values of
$\rho$ or $k$ if to determine one of these values purely model
presentations about another value  are used;

\item
the specific character of the transfer of errors $\delta \rho$
and $\delta k$ to an error $\delta S$ of the calculated spectrum;

\item
all aspects of the influence of the structure of the excited
levels of nucleus on the found parameters $\rho$ and $k$.
\end{itemize}

All these problems become apparent to the full during the
calculation of  the gamma-ray spectra of the radiative capture
of thermal neutrons, measured, for example, by Groshev [11],
with the use of $\rho$ and $k$,  determined from the gamma-ray
intensities in the procedures [8] or [2,3].
The major part of experimental data on the total
gamma-ray spectra of the capture not only of thermal but also
fast neutrons was used to verify such data earlier [12].

The measurement [13] of total gamma-spectra in the isotopes
$^{118,120} $Sn makes it possible to carry out the same analysis
for the spherical nuclei from the region of minimum of neutron
strength function for the s-neutrons.  And to thus to test
obtained and represented into [3] the values $\rho$ and  $k$.

\section{The comparison of calculation and experiment}

With the comparison it is necessary to consider the specific
character of the operation of the transfer of errors $\delta \rho$,
$\delta k(E1)$ and $\delta k(M1)$ to an error in the calculated
total gamma-spectrum:
it is characterized by very low coefficients.
(And, obviously, by very large in the opposite case.)

Consequently, calculated spectra, close ones in the quality of
reproduction to the data of experiment can be obtained for the
substantially being differed values $\rho$ and $k$.
And, all the more, this is correct with the presence experimentally
[2] of the established strong dependence of the process of
cascade gamma-decay on the nuclear structure, as the minimum, for
the excitation energies of $E_{ex} <0.5B_n$.

Therefore the comparison of calculation and experiment should be
conducted for the maximum collection of the diverse variants of
functional dependences $\rho$ and $k$ is without fail on the line scale.
It is  most expedient also to perform the comparison of total
gamma-spectra for the spectrum corresponding to the product
of the gamma-quantum intensity on their energy.
The condition $\sum I_\gamma E_\gamma =B_n$ ensures the maximally precise
normalization intensities of the observed gamma-transitionson in average
and the presence of errors of different sign - for different values of
gamma-quantum energies.

Reliable experimental data for $\rho$ and $k$ in the nucleus $^{120}$Sn
for the range of the excitation energies of $E_ {ex} \approx  B_n$
are absent.  Therefore is below for calculating the total gamma-spectrum
in this nucleus of values $\rho$ and $k$ are converted from the results
[3] by the appropriate scaling of these given for the nucleus $^{118}$Sn.

The level densities of both parities and spins 0, 1 and 2 for these
compound nuclei are given in Fig. 1;  the radiative strength functions
of primary E1- and M1-transitions with the maximal coefficients [2]
of an increase in the radiative
strength functions of secondary transitions are given in Fig. 2
respectively. The calculated total gamma-spectra of the capture of
fast neutrons in the $^{117,119}$Sn are compared with the experiment
in Fig. 3 and 4 for few sets of the values of the
level density and radiative strength functions.
Corresponding calculation data for $^{118,120} $Sn are given in Fig. 3,4.

These data are acquired as follows: the level density and radiative
strength functions are extrapolated to the excitation energy
$E_{ex}=B_n+100$ keV.  And then - they were used for calculating the
total gamma-spectra for the spins of the decomposed initial levels
of $J^{\pi}=0^{+}, ~1^{+}$ and for $J^{\pi} =0^{-}, ~1^{-},~2{-}$.
The portion of their contribution to the resulting spectrum was
determined by the fitting of the function of $S^{exp}=k_j S^{cal}(J^{\pi})$.
I.e., with the calculation of total gamma-spectra was considered capture
only by s- and p -neutrons, and the portion of the captures of $k_j$
for each of the possible spins of compound- states was the free parameter.
Obtained  values are given in the table for the minimum $\chi^2$. 
Here one should note, that the minimum $\chi^2$ with the use of
standard model [4,5,14] can be achieved only for negative contribution
of one of the spin states (in the table it substituted to the zero value).
Taking into account great significance $\chi^2$ can be concluded,
that procedure [2] determination $\rho$ and $k$ reproduces experimental
data on the total gamma-spectra is more accurately, than model
presentations of the type [4,5,14]. The existing deviations totally
can be connected only with the inevitable errors of experimental data
for $\rho$ and $k$,  the obtained from the two-step cascades of
capture thermal neutrons into $^ {117} $Sn.
First of all - because of the absence of data according to the
radiative strength functions of primary E1-transitions to the
levels of these nuclei with the excitation energy less than 2-3 MeV.
Or - because of the presence of the strong dependence of the
radiative strength functions of primary transitions from the
structure of the decayed compound-states not only with the small
($E_{ex} < 0.5B_n$) excitation energies, but also for
$0.5B_n <E_{ex}\leq B_n$.\\\\

Table. 
The most probable portion $k_J$ of experimental spectrum, corresponding to the
decay of compound-state with the spin of $J^{\pi}$ with the use of
experimental data [2] and model presentations [4] and [14] for the strength
functions and the level density.\\
 
\begin{tabular}{|c|r|l|r|l|}
\hline
\multicolumn{3}{|c|}{$^{118}$Sn} &\multicolumn{2}{c|}{$^{120}$Sn}\\\hline
$J^{\pi}$& [4,14]  & [2]
& [4,14]  & [2]
\\\hline
$0^{+}$ & 0.16(16) & 0.00(30)& 0.21(86)&  0.18(21)\\
$1^{+}$ & 0.18(34) & 0.39(33)& 0.00(21)&  0.29(16)\\
$0^{-}$ & 0.21(54) & 0.00(30)& 0.12(98)&  0.28(16)\\
$1^{-}$ & 0.21(75) & 0.01(30)& 0.30(18)&  0.07(6)\\
$2^{-}$ & 0.16(91) & 0.53(83)& 0.15(17)&  0.04(6)\\\hline
sum     & 0.92     & 0.93      &  0.78   &  0.86 \\\hline
\end{tabular}\\

The results of the comparison of the spectra, calculated for different
functional dependencies of level density and of the strength functions
of dipole gamma-transitions with the experiment, quite
unambiguously lead to the conclusion, fully coinciding with those obtained
earlier:

\begin{itemize}
   \item 
``smooth" function $\rho=f(E_{ex})$ reproduces the total gamma-spectrum
of the thermal neutron capture noticeably worse, than the stepped
functional dependencies obtained in [2,3];
\item 
Is hence automatic (because of the strong correlation $\rho$ and $k$)
follows the impossibility of the precise description of the experimental
values of $k$ by model presentations of the type [4,5].
\end{itemize}

In particular, taking into account the influence of the structure of the
excited levels on a change in the form of the energy dependence
of radiative strength functions
most likely should be carried out up to the neutron binding energy.
One must not exclude the possibility that the radiative
and neutron strength functions also depend on the structure of neutron
resonances at the excitation energies larger, than $B_n$.

\section{Conclusion}

The comparison of the total gamma-spectra for different functional dependencies
of $\rho$ and $k(E1)+k(M1)$ both on the excitation energy of nucleus and on
the energies corresponding to the primary and secondary gamma-transitions
for the thermal neutrons capture in $^{117,119}$Sn with the experimental data
was carried out.
The comparison showed that model predictions of the non-interacting Fermi gas
level density in these nuclei give worse
correspondence, than the level density from the procedures [2,3].
This conclusion corresponded to the one obtained earlier [12].

Large transfer coefficients of the errors $\delta S$ of total
gamma-spectra to the errors
$\delta \rho$ and $\delta(k(E1)+k(M1))$ directly follow from the comparison
of the data in Figs. 1,2 and 3,4.  This circumstance confirms the conclusion [1],
that the measurement of such spectra,  for example in the procedure [8],
requires accuracy on $\sim 2$ orders larger, than in the procedure [2,3].
And it limits the possibilities of the independent checking of different
sets of $\rho$ and $k$, both of 
model determined ones and of experimentally obtained ones.
The use of total gamma-spectra for their testing necessary requires
the comparison of different variants of such data.

And even total reproduction of the experimental total gamma-spectrum by
calculation with a certain set of $\rho$ and $k$ is not the proof of the
correspondence of these values to the real parameters of nucleus.
However, explicit nonconformity is a quite single-valued proof of the presence
of larger or smaller systematic deviation for them with the experimental one.
\newpage

\begin{flushleft}
{\large\bf References}\end{flushleft}\begin{flushleft}
\begin{tabular}{r@{ }p{5.65in}} 
$[1]$ & A.M. Sukhovoj, V.A. Khitrov,
      Physics of Paricl. and Nuclei 37(6) (2006) 899.\\
      & A.M.  Sukhovoj, V.A.  Khitrov, JINR preprint E3-2005-196, Dubna,
      2005.\\
      & http://www1.jinr.ru/Preprints/Preprints-index.html\\
$[2]$ & A.M.  Sukhovoj, V.A.  Khitrov, Physics of Paricl. and Nuclei,
 36(4) (2005) pp. 359-377.\\
      & http://www1.jinr.ru/Pepan/Pepan-index.html (in Russian)\\
$[3]$ & E.V.  Vasilieva, A.M.  Sukhovoj, V.A.  Khitrov, Phys.  At. Nucl. 
64(2) (2001) 153, nucl-ex/0110017\\
$[4]$ & S.G. Kadmenskij, V.P. Markushev, V.I. Furman, Sov. J. Nucl. Phys. 37 (1983) 165.\\
$[5]$ & P. Axel,  Phys. Rev. 1962. 126. $N^o$ 2. P. 671.\\
$[6]$ & V.G. Soloviev, Phys. of Elementary Particles and Atomic Nuclei, 3(4) (1972) 770.\\
$[7]$ & B.V. Zhuravlev, Bull. Rus. Acad. Sci. Phys. 63 (1999) 123.\\
$[8]$ & G.A.  Bartholomew et al., Advances in nuclear physics 7 (1973) 229.\\
      & A. Schiller et al., Nucl.  Instrum. Methods Phys.  Res. A447 (2000) 498.\\
$[9]$ & I.A. Lomachenkov, W.I. Furman, JINR, P4-85-466, Dubna, 1985\\
$[10]$ &  O.T.  Grudzevich,  Phys.  At. Nucl.  62 (1999) 192.\\
$[11]$ & L.V. Groshev et al., Atlac thermal neutron capture gamma-rays spectra,
 Moscow, 1958.\\
$[12]$ & A.M. Sukhovoj, V.A. Khitrov and  E.P. Grigor'ev,
 INDC(CCP)-432, Vienna 115 (2002).\\
  & V.A. Khitrov, A.M. Sukhovoj, Pham Dinh Khang, Vuong Huu Tan, Nguyen Xuan Hai,
XIII International Seminar on Interaction of Neutrons with Nuclei,  Dubna, 22-25 May 2005,
E3-2006-7, Dubna, 2006, p. 64, nucl-ex/0508008\\
$[13]$ & J.Nishyama, M.Igashira, T. Ohsaki, G.N.Kim, W.C.Chung, T.I.Ro,
Twelfth International Symposium on Capture Gamma-Ray Spectroscopy
   and Related Topics, Notre Dame, September 4-9, 2005, World Scientific,
   Ed. A. Woehr, A. Aprahamian, p. 579.\\
   &  J. Nishyama, T.I. Ro, M. Igashira, W.C. Chung, G. Kim, T. Ohsaki,
S. Lee, T. Katabuchi, International Conference on Nuclear Data for Science and
Technology 2007, Nice, April 2007, to be published.\\
$[14]$ &W. Dilg, W. Schantl, H. Vonach, M. Uhl, Nucl. Phys. A217 (1973) 269.\\
\end{tabular} 
\end{flushleft}


\begin{figure}
\vspace{-3cm}
\begin{center}
\leavevmode
\epsfxsize=15.5cm
\epsfbox{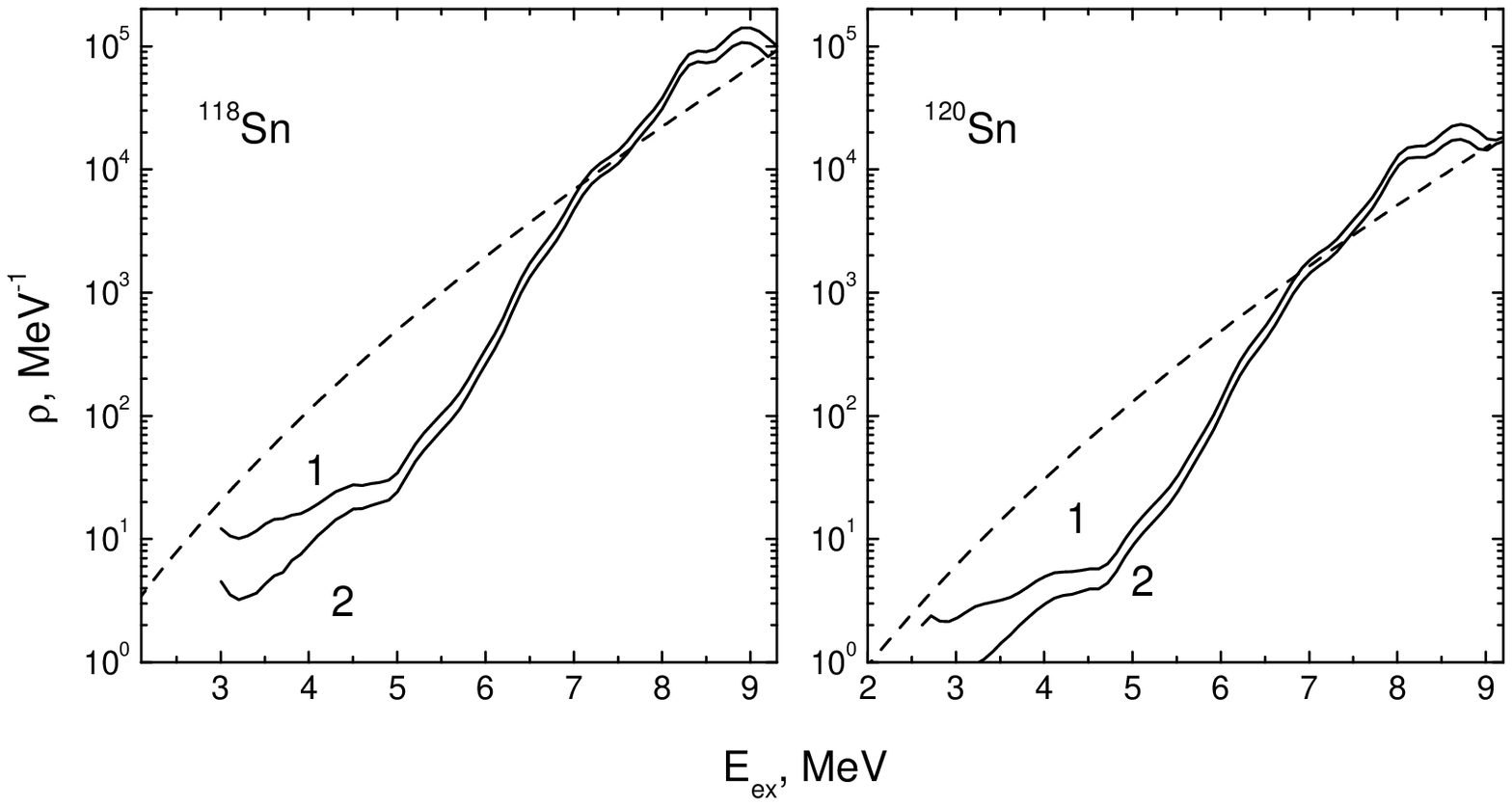}
\end{center}
\hspace{-0.8cm}
\vspace{-14cm}

{\bf Fig.~1.}
The level densities of both parities and spins
0, 1 and 2 for
compound nuclei $^{118,120}$Sn. Line 1 presented the used in calculations
values of the level density with spins J=(0-2)$^{+}$, line 2 - the
same for negative parity only.
  The dash line represents predictions of the model [14].
\end{figure}

\begin{figure}
\vspace{-4cm}
\begin{center}
\leavevmode
\epsfxsize=15.5cm
\epsfbox{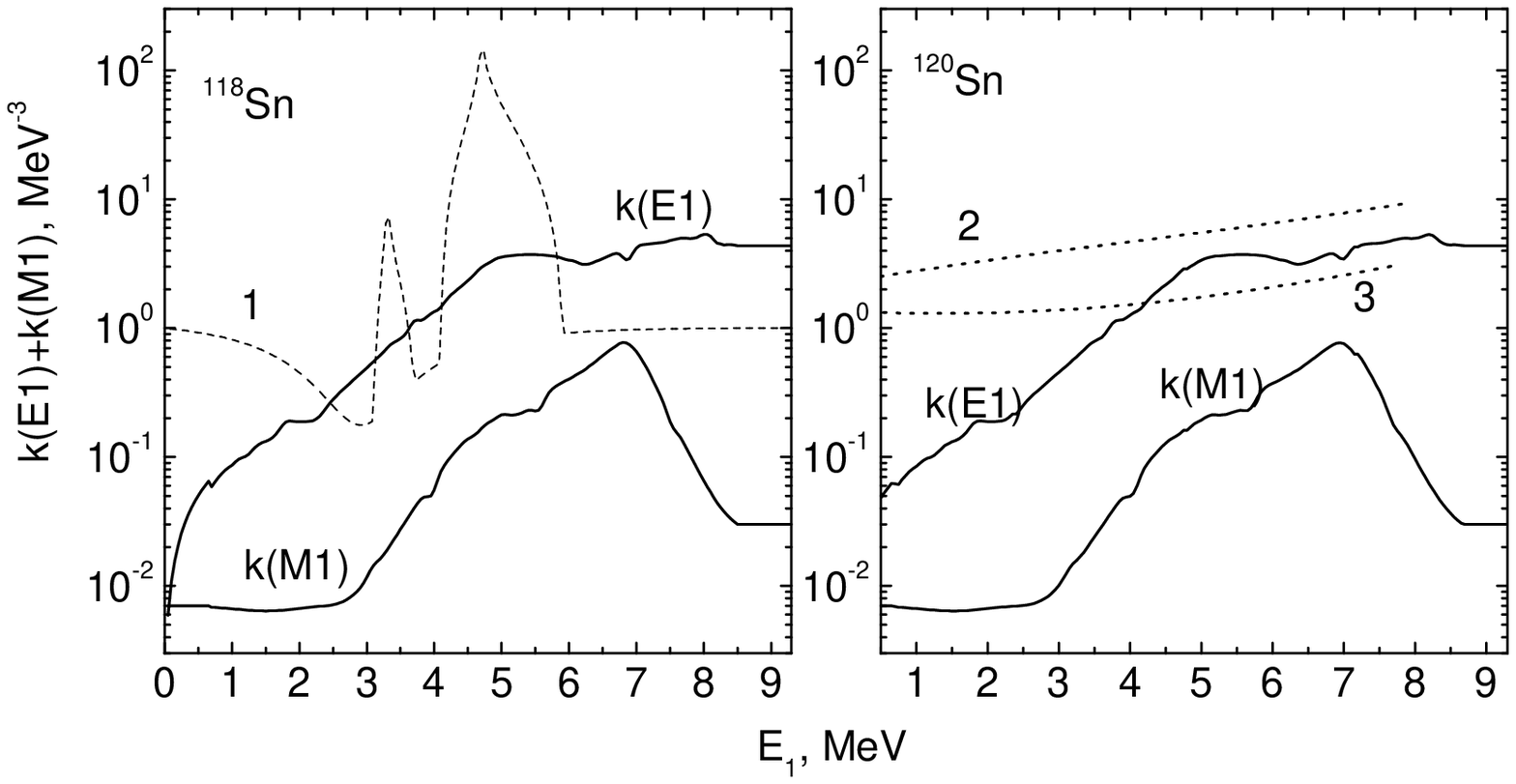}
\end{center}
\hspace{-0.8cm}

\vspace{-14cm}

{\bf Fig.~2.} Solid lines presented the radiative strength functions of primary
E1- and  M1-transitions (multipleted on $10^{9}$).
Line 1- maximum increasing of the radiative strength functions of secondary
transitions to the levels with the energy of $E_{ex}=B_n-E_1$.

Line 2 and 3 represent predictions of the model [5] and [4] in the sum with
$k(M1)$=const.  For both models is used the ratio $k(M1)/k(E1)=0.2$
for of $E_1 =6.5$ MeV.
\end{figure}
\newpage
\begin{figure}
\vspace{-3cm}
\begin{center}
\leavevmode
\epsfxsize=15.5cm
\epsfbox{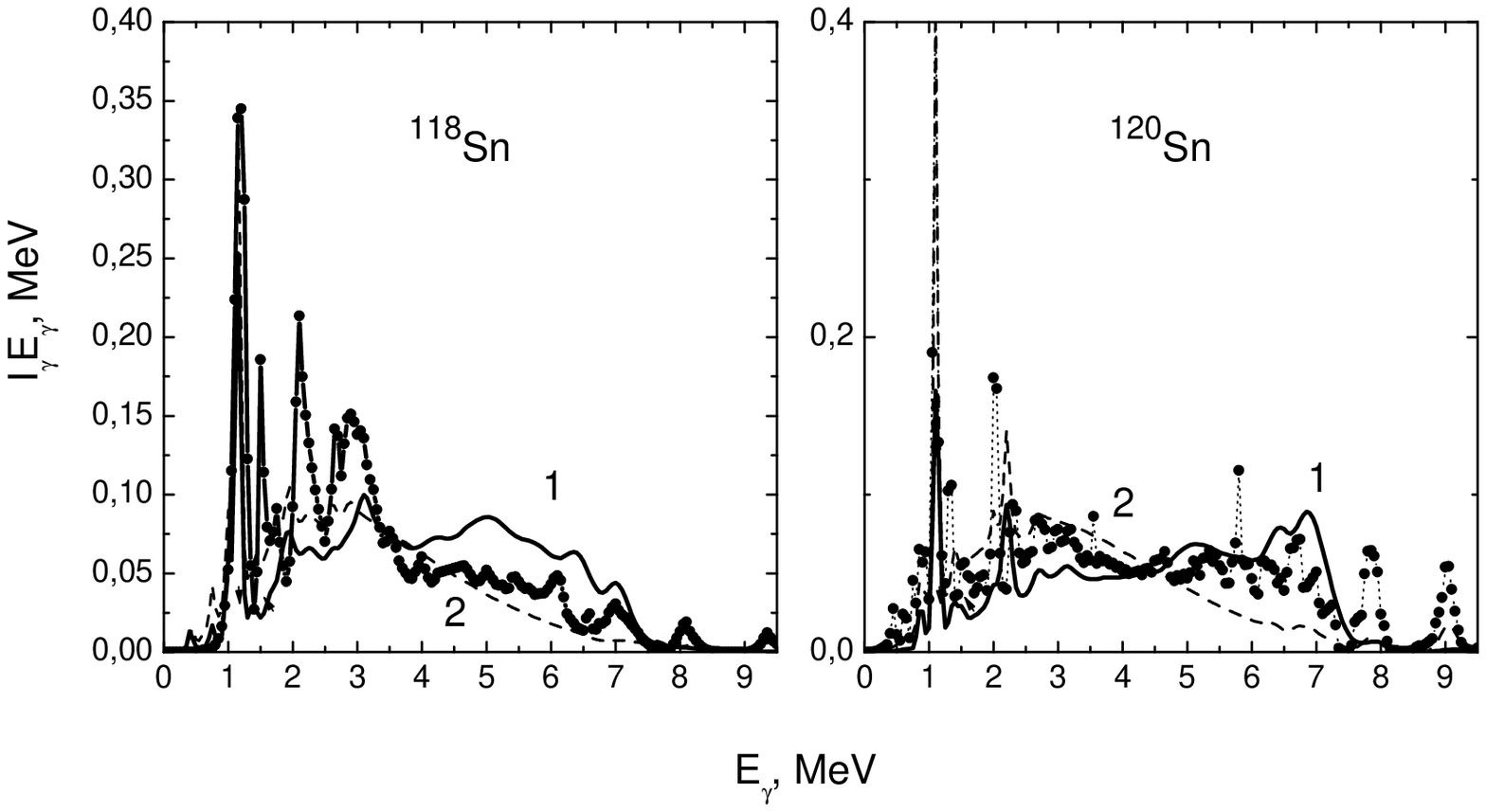}
\end{center}
\hspace{-0.8cm}\vspace{-14cm}

{\bf Fig.~3.} The experimental  (points+line) total spectra of
$\gamma$-radiation following fast neutron capture for the
 $^{117,119}$Sn targets.
Lines 1 represent results of calculation
using data of Ref. [2], line 2 - from [4,14], corresponding.
\end{figure}
\begin{figure}

\vspace{-5cm}
\begin{center}
\leavevmode
\epsfxsize=17.cm
\epsfbox{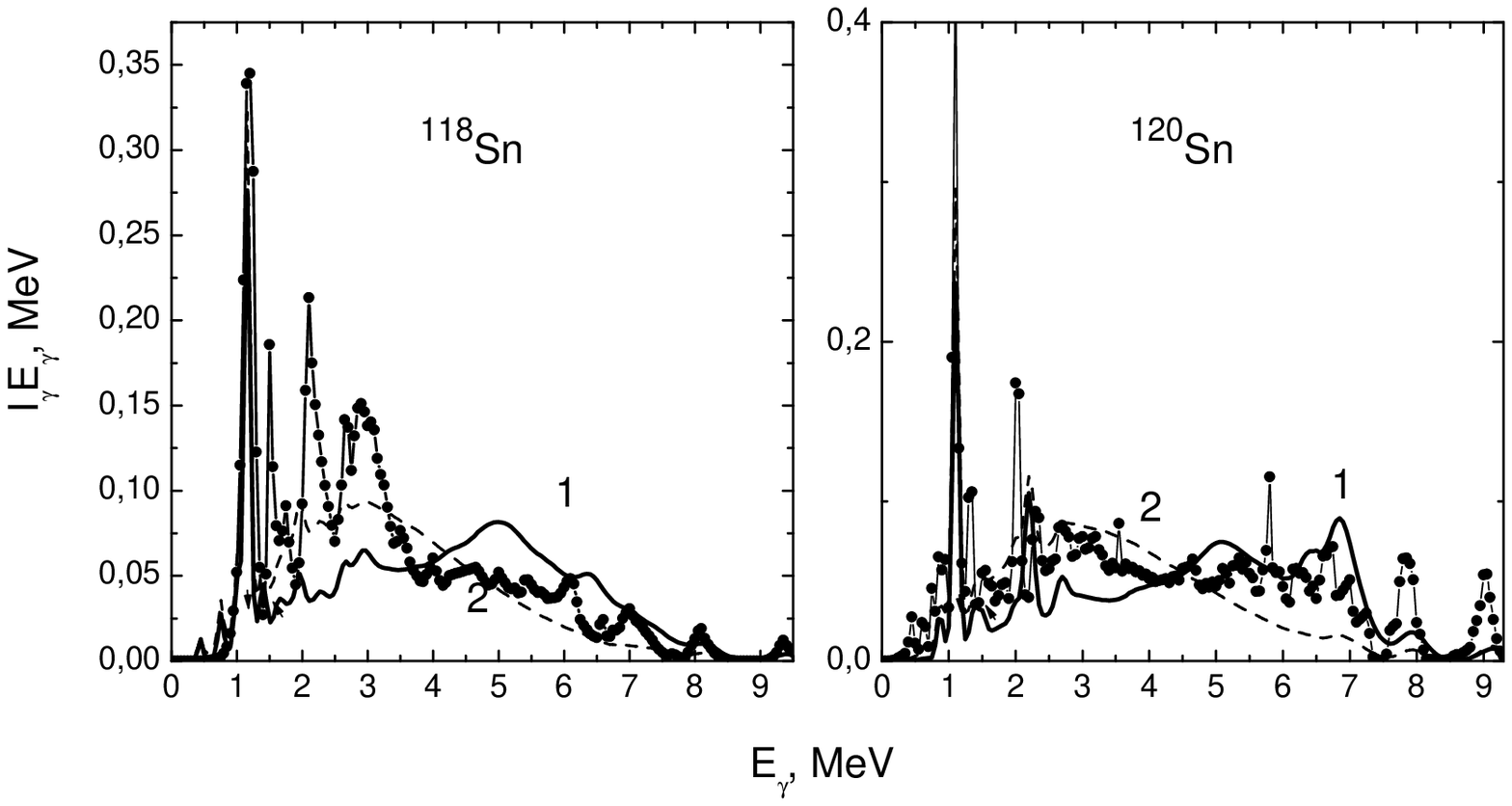}
\end{center}

\vspace{-14cm}

{\bf Fig.~4.} The same, as on Fig. 3.
Lines 1 represent results of calculation
using data of Ref. [3], lines 2 - from [5,14], corresponding. 
\end{figure}

\end{document}